\documentclass{mem}
\usepackage{natbib}\usepackage{txfonts}\usepackage{balance}
\usepackage{graphicx}
\usepackage[a4paper]{hyperref}
\idline{75}{282}
\begin{document}
\def\teff{$T\rm_{eff }$}
\def\kms{$\mathrm {km s}^{-1}$}

\title{
Gamma-Ray Flares from Mrk421 in 2008 observed with the ARGO-YBJ
experiment
}

   \subtitle{}

\author{
G. \,Di Sciascio on behalf of the ARGO-YBJ Collaboration
          }

  \offprints{G. Di Sciascio}

\institute{INFN -- Sez. di Roma Tor Vergata, Viale della Ricerca
Scientifica 1, I-00133 Roma, Italy\\ \email{disciascio@roma2.infn.it} }

\authorrunning{Di Sciascio }

\titlerunning{Mrk421 with ARGO-YBJ}

\abstract{In 2008, the blazar Mrk421 entered in a very active phase and
was one of the brightest sources in the sky at TeV energies, showing
strong and frequent flaring. We searched for $\gamma$-ray emission at
energies E$>$0.8 TeV during the whole 2008 with the ARGO-YBJ experiment,
a full coverage air shower detector located at Yangbajing (4300 m a.s.l.,
Tibet, P.R. China). The observed signal is not constant and in
correlation with X-ray measurements. The average emission, during the
active period of the source, was about twice the Crab Nebula level, with
an integral flux of (4.9$\pm 2.0$)$\times$10$^{-11}$ $\gamma$ cm$^{-2}$
s$^{-1}$ for E$_{\gamma}$$>$1 TeV.

This paper concentrates on 2008 June when the Mrk421 flaring activity has
been studied from optical to 100 MeV gamma rays, and only partially up to
TeV energies, since the moonlight hampered the Cherenkov telescope
observations after 8 June. Our data complete these observations, with the
detection of a second flare of intensity of about 7 Crab units on June
11-13, with a statistical significance of 4.2 standard deviations. The
observed flux is consistent with a prediction made in the framework of
the Synchrotron Self-Compton model, in which the flare is caused by a
rapid acceleration of leptons in the jet.  } \maketitle{}

%=============================================================
\section{Introduction}
%=============================================================

Mrk421 ($z$=0.031) is the brightest BL Lac object and the first
extragalactic source discovered at TeV energies \citep{Pun92}, where
dramatic variability has been observed with flux increasing by more than
a factor 50 in about one hour \citep{Gai96}. Like most blazars, its
spectral energy distribution (SED) shows two smooth broadband components,
the first one peaking in the soft to medium X-ray range and the second
one extending to the GeV/TeV energies. X-rays are generally attributed to
synchrotron radiation from high energy electrons, while the origin of the
$\gamma$-ray emission is more uncertain. Possibilities include inverse
Compton scattering of synchrotron (Synchrotron Self-Compton, SSC) or
ambient photons (external Compton, EC) off a single electron population
\citep{Sam96,Ghi98,Fos08}. Nevertheless, an alternative hadronic model
($\gamma$-rays from proton synchrotron \citep{Muc03}) for Mrk421 is not
ruled out yet.

Multiwavelength observations are the key for understanding the blazar
phenomenon. Indeed, it is generally true that hadronic models are in
trouble to reproduce the observed highly correlated X-ray/TeV
variability, that strongly supports the SSC models
\citep{Fos08,Fid08,Wag08}.

The ARGO-YBJ experiment is presently the only wide-field of view
$\gamma$-ray telescope able to detect AGN TeV flaring activity on a few
day period. In this paper we report on the monitoring of Mrk421 performed
with ARGO-YBJ in 2008. We will give special attention to the 2008 June
flaring activity because an extraordinary set of simultaneous
measurements covering 12 decades of energy, from optical to TeV gamma
rays, is available \citep{Don09}.

%=============================================================
\section{The ARGO-YBJ experiment}
%=============================================================

The ARGO-YBJ detector, located at the YangBaJing Cosmic Ray Laboratory
(4300 m a.s.l., Tibet, P.R. China), is the only experiment exploiting the
\emph{full coverage} approach at very high altitude. The detector is
constituted by a central carpet $\sim$74$\times$78 m$^2$, made of a
single layer of Resistive Plate Chambers (RPCs) with $\sim$92$\%$ of
active area, enclosed by a partially instrumented guard ring that extends
the detector surface up to $\sim$100$\times$110 m$^2$. The apparatus has
a modular structure, the basic data acquisition element being a cluster
(5.72$\times$7.64 m$^2$), divided into 12 RPCs (2.80$\times$1.25 m$^2$
each). Each chamber is read by 80 strips of 7$\times$62 cm$^2$ (the space
pixel), logically organized in 10 independent pads of 56$\times$62 cm$^2$
representing the time pixel of the detector. The RPCs are operated in
streamer mode with a standard gas mixture (Argon 15\%, Isobutane 10\%,
TetraFluoroEthane 75\%), the High Voltage settled at 7.2 kV ensures an
overall efficiency of about 96\% \citep{Aie06}. The full detector is
composed of 153 clusters for a total active surface of $\sim$6700 m$^2$.
All events giving a number of fired pads N$_{pad}\ge$ N$_{trig}$ in the
central carpet within a time window of 420 ns are recorded. The spatial
coordinates and the time of any fired pad are used to reconstruct the
position of the shower core and the arrival direction of the primary, as
described in \citep{DiS07,DiS08}.

Since 2007 November the full detector is in stable data taking at the
multiplicity trigger threshold N$_{trig}\geq$20 with a duty cycle $\sim
90\%$: the trigger rate is about 3.6 kHz.

%=============================================================
\section{Data analysis}
%=============================================================

The dataset for the analysis of Mrk421 presented in this paper contains
all showers with N$_{pad} \geq$40 and zenith angle $\theta<$40$^{\circ}$.

A 20$^{\circ}\times$ 20$^{\circ}$ sky map in celestial coordinates (right
ascension and declination) with 0.1$^{\circ}\times$0.1$^{\circ}$ bin
size, centered on the source location, is filled with the detected
events. The background is evaluated with the \emph{time swapping} method
\citep{Ale92}. N \emph{"fake"} events are generated for each detected
one, by replacing the measured arrival time with new ones. These times
are randomly selected within a 3 hours wide buffer of recorded data.
Swapping the time means swapping the right ascension, keeping unchanged
the declination. A new sky map (background map) is built by using 10 such
fake events for each real one, so that the statistical error on the
background can be kept small enough.

To maximize the signal to noise ratio, the bins are then grouped over a
circular area of radius $\psi$, i.e. every bin is filled with the content
of all the surrounding bins whose center is closer than $\psi$ from its
own center. When the Point Spread Function of the detector is a Gaussian
with r.m.s. $\sigma$, the opening angle $\psi$ containing 71.5$\%$ of the
events maximizes the signal to background ratio for a point source with a
uniform background, and it is equal to 1.58$\cdot\sigma$. The values of
$\psi$ are 1.9$^{\circ}$, 0.9$^{\circ}$ and 0.5$^{\circ}$ respectively
for N$_{pad}\geq$40, 100 and 300, in agreement with MonteCarlo
simulations \citep{Ver09a}.

Finally, the integrated background map is subtracted from the
corresponding integrated event map, thus obtaining the "source map". For
each bin of this latter map, the significance with respect to the
background is calculated.

This analysis procedure has been tested with the Crab Nebula, the
standard candle for VHE astronomy. At the Yangbajing latitude the Crab
culminates at zenith angle $\theta_{culm}$ = 8.1$^{\circ}$ and it is
observable every day for 5.8 hours with a zenith angle $\theta
<$40$^{\circ}$. The Crab Nebula has been observed from 2007 November to
2009 March, for a total of 424 days on-source, obtaining a signal with a
statistical significance of 7 standard deviations for N$_{pad}\ge$40. The
corresponding photon median energy is 1.1 TeV. In this analysis no events
selection or $\gamma$/hadron discrimination algorithm has been applied.
The average number of gamma rays detected per day in the observational
window centered on the source position is 155$\pm$25.

\begin{figure}[]
\resizebox{\hsize}{!}{\includegraphics[clip=true]{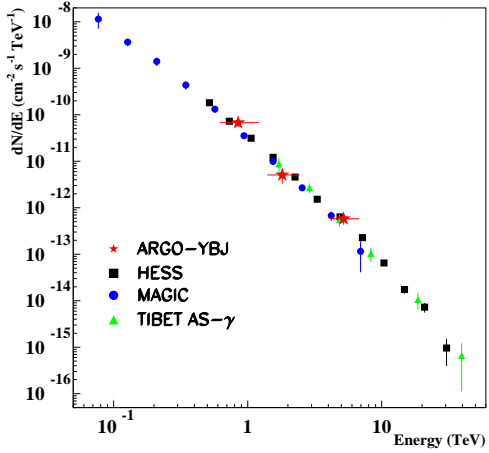}}
  \caption{\footnotesize The Crab Nebula energy spectrum measured in 2008 by ARGO-YBJ compared with the results of some other detectors. }
\label{crab}
\end{figure}

By assuming a differential spectrum dN/dE = K$\cdot E^{-\alpha}$, we
simulated a source in the sky following the diurnal path of the Crab, and
evaluated the number of events expected in the 3 N$_{pad}$ bins 40-99,
100-299 and $\ge$300, for different values of K and $\alpha$
(10$^{-11}$$<$K$<$10$^{-10}$ photons cm$^{-2}$ s$^{-1}$ TeV$^{-1}$ and
1.5$< \alpha <$3.5). The multiplicity bins correspond respectively to
median energies 0.8, 1.8 and 5.2 TeV. Comparing the expected values with
the observed ones we obtained the spectrum that best fits to data
\begin{equation}
dN/dE = (3.7\pm 0.8) \times 10^{-11} E^{-2.67\pm 0.25}
\end{equation}
in fair agreement with other experiments (see Fig. \ref{crab}).
%

%=============================================================
\section{Mrk421 analysis}
%=============================================================

The same analysis has been performed for Mrk421. This source culminates
at the ARGO-YBJ location at zenith angle $\theta_{culm} = 8.1^{\circ}$,
and it is observable every day for 6.38 hours with a zenith angle $\theta
<$ 40$^{\circ}$.

Here the Mrk421 data collected from 2007 day 311 to 2008 day 366 are
presented. Using the same method adopted for the Crab Nebula, we
evaluated the Mrk421 spectrum from 2008 day 41 to day 180, where the
X-ray flux showed the most intense flares. In this period (755
observation hours) the observed signal had a statistical significance of
6.1 standard deviations.
\begin{figure}[]
\resizebox{\hsize}{!}{\includegraphics[clip=true]{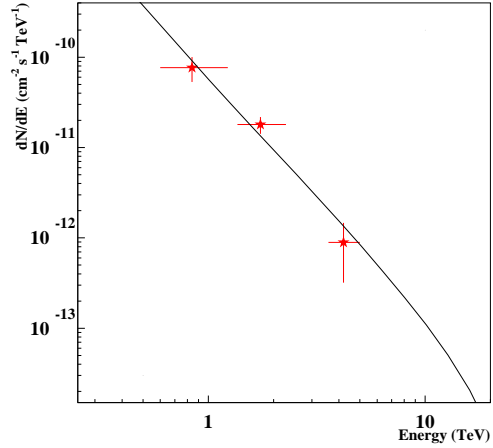}}
 \caption{\footnotesize The Mrk421 energy spectrum measured by ARGO-YBJ from 2008 day 41 to 180, when the source was in active state.
The line represents the best fit to the data.}
\label{mrk1}
\end{figure}
The differential spectrum (photons cm$^{-2}$ s$^{-1}$ TeV$^{-1}$) that
best fits to data is (see Fig. \ref{mrk1})
\begin{equation}
dN/dE  = (7.46\pm 1.70) \times 10^{-11} E^{-2.51\pm 0.29} e^{-\tau(E)}
\end{equation}
where the exponential factor e$^{-\tau(E)}$ takes into account the
absorption of gamma rays in the Extragalactic Background Light, with
$\tau(E)$ given in \citep{Pri05}. The integral flux above 1 TeV is
(4.9$\pm2.0$)$\times$10$^{-11}$ photons cm$^{-2}$ s$^{-1}$, about twice
the Crab Nebula flux. The median energies corresponding to the 3
N$_{pad}$ bins 40-99, 100-299 and $\ge$300 are respectively: E =
0.84$^{+0.39}_{-0.24}$ , 1.74$^{+0.54}_{-0.38}$ and 4.2$^{+0.79}_{-0.62}$
TeV.

\begin{figure}[]
\resizebox{\hsize}{!}{\includegraphics[clip=true]{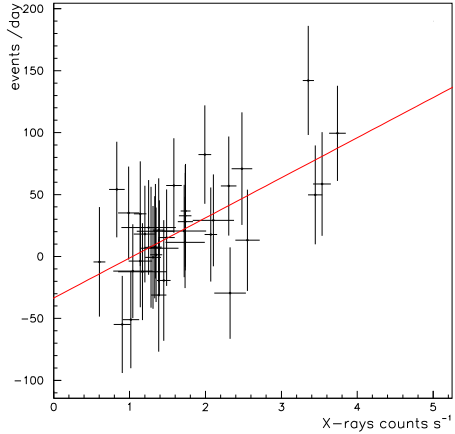}}
 \caption{\footnotesize Plot of the RXTE/ASM X-ray and ARGO-YBJ TeV gamma-ray count rates of Mrk421. The line shows the best linear fit.
} \label{correl-gamma-xrays}
\end{figure}

The observed gamma ray rate appears to be correlated with the X-ray rate
measured by the All Sky Monitor detector aboard the RXTE satellite in the
1.5-12 keV energy range \citep{RXTE}, as can be seen in Fig.
\ref{correl-gamma-xrays} where the X-ray and ARGO-YBJ TeV gamma-ray count
rates are shown for all the simultaneous measurements in 2008. The excess
events observed by ARGO-YBJ, averaged over 10 days, refer to showers with
N$_{pad}\ge$100. A positive correlation between the rates seems apparent:
the correlation coefficient is 0.64.

Since Mrk421 is known to vary on different time scales, the source has
been studied during 1, 10 and 30 days \citep{Ver09b}. For this analysis
we have considered the data taken in the period 2007 day 311 - 2009 day
89. No excess has been observed on a daily scale. Concerning the 10
scale, we observed an excess at 4.6 s.d. in the time interval 2008 days
161 - 170, during a strong X-ray flare. Looking for 30 days excesses, the
search has been carried out by shifting the 30 days intervals in steps of
10 days. We found several excesses from Mrk421 with significances between
4 and 5 s.d., in particular in the intervals: 2008 days 1 - 30, 71 - 100,
81 - 110, 91 - 120, 141 - 170, when several X-ray flares have been
observed.

\subsection{The June 2008 flares}

A set of simultaneous measurements covering 12 decades of energy, from
optical to TeV gamma rays, was performed during the strong flaring
activity in the first half of June 2008 by different detectors: WEBT
(optical R-band), SWIFT (UV, soft and hard X-rays), RXTE/ASM (soft
X-rays), AGILE (hard X-rays and gamma rays) and the Cherenkov telescopes
VERITAS and MAGIC (VHE gamma rays) \citep{Don09}.

In this period two flaring episodes were reported, the first one on June
3 - 8, observed from optical to TeV gamma rays, the second one, larger
and harder, on June 9 - 15, observed from optical to 100 MeV gamma rays.
Using this multi-frequency data, in \citep{Don09} the authors derived the
SED for June 6, that shows the typical two humps shape, in agreement with
the SSC model. According to the authors, the second hump intensity (that
reached a flux of about 3.5 Crab units at energy E $>$400 GeV) seems to
indicate that the variability is due to the hardening/softening of the
electron spectrum, and not to the increase/decrease of the electron
density. Their model predicts for the second flare a VHE flux about a
factor 2 larger with respect to the first one. Unfortunately there were
no VHE data included in their multi-wavelength analysis after June 8
because the moonlight hampered the Cherenkov telescopes measurements.

The VHE observation was actually made by the ARGO-YBJ experiment, that
since 2007 November is performing a continuous monitoring of Mrk421.
\begin{figure}[]
\resizebox{\hsize}{!}{\includegraphics[clip=true]{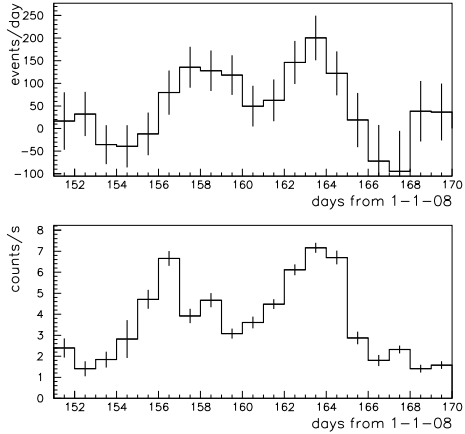}}
 \caption{\footnotesize Upper panel: rate of events with  N$_{pad} \ge$100 observed by ARGO-YBJ as a function of time
from May 31 00:00 UT to June 19 00:00 UT.
Each bin contains the rate averaged over the 3 days interval centered
on that bin. Lower panel: daily counting rate of RXTE/ASM.}
\label{gamma-xrays-june08}
\end{figure}
Fig. \ref{gamma-xrays-june08} shows the rate of events with N$_{pad}
\ge$100 observed by ARGO-YBJ during the first half of June, averaged over
3 days, compared with the X-ray flux measured by RXTE/ASM.

\begin{figure}[]
\resizebox{\hsize}{!}{\includegraphics[clip=true]{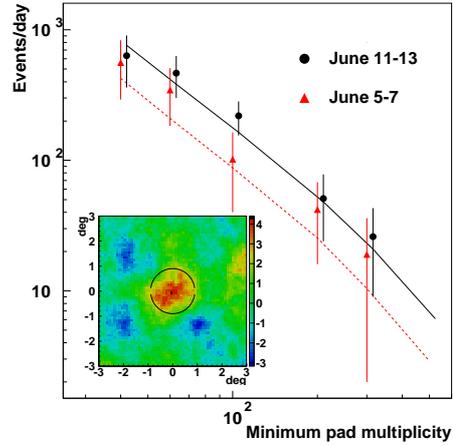}}
 \caption{\footnotesize Event rates observed by ARGO-YBJ as a function of the minimum
pad multiplicity on June 5-7 and June 11-13 (respectively triangles
and circles) compared with expected rates according to the Donnarumma et al. model
for the same two periods (respectively dashed and solid lines).
The inset represents the sky map around the Mrk421 position on
June 11-13, obtained for events with N$_{pad}$$\ge$100. The color scale
shows the significance of the signal in standard deviations.
The circle represents the observational window of radius 0.9$^{\circ}$. }
\label{mrk421-june08}
\end{figure}

A second flare has been detected with a statistical significance of 3.2
standard deviations during the interval 11-13 June. The significance
increases to 4.2 s.d. with a suitable events selection \citep{Ver09a}.
Fig. \ref{mrk421-june08} shows the 6$^{\circ}\times$6$^{\circ}$ sky map
around the source position in these 3 days, after applying the event
selection. For every 0.1$^{\circ}\times$0.1$^{\circ}$ bin, the map gives
the value of the statistical significance of the excess of events inside
the circular window of radius 0.9$^{\circ}$ centered on that bin.
Donnarumma et al. evaluate a theoretical SED curve for the first flare
fitting the observations made on June 6 from optical up to VHE energies.
During the second flare, they reported a higher photon flux from soft
X-rays to 100 MeV gamma rays. From these data they predict a flux at
E$>$1 TeV of 1.45$\cdot$10$^{-10}$ photons cm$^{-2}$ s$^{-1}$
corresponding to about 7 Crab units, and they model a SED curve with the
Inverse Compton hump slightly shifted towards higher energies.

Fig. \ref{mrk421-june08} shows the event rate observed by ARGO-YBJ (in
18.2 hours of measurement) compared with the rate expected from a source
spectrum given by the theoretical SED, for any N$_{pad}$ interval. The
agreement is good. A similar analysis is made for the first flare,
integrating our data on June 5-7 (17.9 hours of measurement). The
observed signal has a significance of $\sim$2 standard deviations,
increased to 3 using the data selection. The event rate obtained as a
function of the minimum pad multiplicity, is consistent with the one
predicted by the SED of Donnarumma et al. for June 6.

\begin{figure}[]
\resizebox{\hsize}{!}{\includegraphics[clip=true]{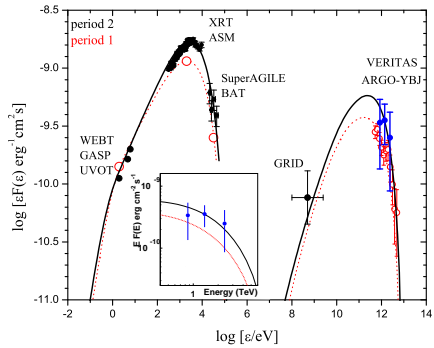}}
 \caption{\footnotesize SED measured by ARGO-YBJ
on June 11-13 (large blue filled circles) together with the data of
other experiments, obtained during the first flare (open circles),
and the second one (filled circles).
The curves represent the SEDs modeled in \citet{Don09} for the
first flare (dashed line) and the second one (solid line).
The inset shows a zoom on the ARGO-YBJ data. }
\label{mrk421-sed}
\end{figure}

Finally we estimate the energy spectrum for the second flare (the
marginal significance of the first flare doesn't allow the spectrum
evaluation). Assuming a source spectrum given by the theoretical SED, the
median energy of the events in the 3 pad multiplicity bins (40 - 99, 100
- 299 and $\ge$300) are respectively 0.9, 1.4 and 2.4 TeV. Fig.
\ref{mrk421-sed} gives the measured SED for the 3 energy points
((3.38$\pm$2.03)$\cdot 10^{-10}$, (3.55$\pm$1.37)$\cdot 10^{-10}$ and
(2.49$\pm$1.63)$\cdot 10^{-10}$ erg$^{-1}$cm$^{2}$s), together with all
the measurements in the optical-TeV range, and the theoretical SED for
the two flares. The measurements appear in fair agreement with the
expected emission.

%=============================================================
\section{Conclusions}
%=============================================================

In summary, Mrk421 has been continuously monitored with ARGO-YBJ during
2008, showing a VHE flux twice the Crab Nebula level from day 41 to 180,
when the source was in active phase, and decreasing afterwards.

ARGO-YBJ observed a flare on 5-7 June, with a flux about 3.5 Crab units,
in agreement with the VERITAS observation. A second flare has been
discovered with a statistical significance of 4.2 s.d., during the
interval 11-13 June, with a flux about 7 Crab units.

ARGO-YBJ measured the spectra of Mrk421 above 0.8 TeV during the second
flare completing a multiwavelength campaign from optical to TeV energies.

For the first time an air shower array was able to detect gamma-ray
flaring activity at sub-TeV energies on a few days period.

\begin{acknowledgements}
We are grateful to the authors of \citet{Don09}, in particular to Marco
Tavani and the AGILE team, for helpful discussions and for providing us
full details on the Mrk421 broadband data relative to the published
analysis.
\end{acknowledgements}

\bibliographystyle{aa}

\end{document}